\documentclass[twocolumn,superscriptaddress,showpacs,aps,amsmath,amssymb]{revtex4}
\usepackage{bm}
\usepackage{graphicx}

\newcommand{\B}{\text{\scriptsize res}}
\newcommand{\s}{\text{\scriptsize sys}}

\newcommand{\T}{{\rm total}}

\newcommand{\nl}{\nonumber \\}

\newcommand{\ep}{\epsilon}
\newcommand{\w}{\omega}

\newcommand{\be}{\begin{equation}}
\newcommand{\ee}{\end{equation}}
\newcommand{\bea}{\begin{eqnarray}}
\newcommand{\eea}{\end{eqnarray}}
\newcommand{\bsube}{\begin{subequations}}
\newcommand{\esube}{\end{subequations}}

\newcommand{\comments}[1]{}

\begin{document}

\title{Time-Dependent Transport through Quantum-Impurity Systems with Kondo Resonance}

\author{YongXi~Cheng}
\affiliation{Department of Physics, Renmin University of China, Beijing
100872, China }

 \author{ZhenHua~Li}
\affiliation{Department of Physics, Renmin University of China, Beijing
100872, China }
\affiliation{ Beijing Computational Science Research Center, Beijing
100084, China }

\author{WenJie~Hou}
\affiliation{Department of Physics, Renmin University of China, Beijing
100872, China }

\author{JianHua~Wei}\email{wjh@ruc.edu.cn}
\affiliation{Department of Physics, Renmin University of China,
Beijing 100872, China }

\author{YiJing~Yan}
\affiliation{Hefei National Laboratory for Physical Sciences at the Microscale, University of Science and Technology of China, Hefei, Anhui 230026}
\affiliation{China and Department of Chemistry, Hong Kong University of Science and Technology, Kowloon, Hong Kong}

\date{}

\begin{abstract}

We investigate the time-dependent transport properties of single and double quantum-impurity systems based on the hierarchical equations of motion (HEOM) approach. In the Kondo regime, the dynamical current in both cases is found oscillating due to the temporal coherence of electrons tunneling through the device, which shares the same mechanism as the single-level resonance without \emph{e-e} interactions but shows some different characteristics. For single quantum-impurity systems, the temperature $T$ plays an inhibitory action to the oscillations of dynamic current through its suppression to the Kondo effects. The amplitude of the current oscillations is attenuated by the \emph{e-e} interaction $U$ in the Kondo regime. The frequency of the current oscillation is found almost independent of $T$ and $U$. For parallel-coupling double quantum-impurity systems, the oscillation of the current shows similar behaviors to the single one, but with two-to-three times larger amplitudes. At the limit of small inter-impurity coupling the oscillation of the current exhibits enhanced characters while it is weakened at the other limit.

\end{abstract}


\maketitle

\section{Introduction}
Quantum-impurity systems are of great interests because of their fundamental physics as well as potential applications as possible quantum-computing devices \cite{2001prb201311}, such as single-molecule magnets \cite{2012prb104427,2012nature938} and bulk Kondo insulators \cite{2012apl013505}. The investigation of transport in quantum-impurity systems are of great importance to understand the phase coherence of electrons in nanodevices \cite{2005prl196801}. Although the spatial coherence of the electron wave functions in quantum-impurity systems has been widely studied theoretically, the temporal phase coherence has not yet. The reason mainly comes from the difficulty to deal with the memory effects in the time domain, especially when electron-electron (\emph{e-e}) interaction can not be treated perturbatively. Therefore, the \emph{e-e} interactions in most of the studies in literatures on the dynamical current in quantum-impurity systems are either totally ignored or treated on the level of mean field. Some of them are summarized as follows.

 N. S. Wingreen\emph{ et al.} investigate the time-dependent current of the mesoscopic structure coupling a double-barrier relatively early \cite{1993prb11}. A formulation of the time-dependent current is presented in terms of Keldysh Green function. They find that the distinct oscillation of the time-dependent current occurs when a rectangular bias pulse being applied on the device. The reason is attributed to the temporal coherence of electrons tunneling through the resonant level in response to the abrupt change of bias. Then, A theory has been developed by Kohn-Sham equation rely on the so called wide-band limit (WBL), which assumes the band of the leads having no energy dependent features. Although this approach has been widely adopted in mesoscopic physics, it fails to describe leads with a finite bandwidth or with the energy-dependent density-of-states. In Ref.\cite{2005prb075317}, Yu Zhu \emph{et al.} directly solve the Green functions in the time domain beyond the WBL approximation to investigate the dynamical current through a molecular. By applying a bias voltage on device leads, the numerical $I(t)$ is solved by the time domain decomposition(TDD) method. In order to deal with the phase memory, upper time and cutoff are adopted in the TDD for an open system \cite{2005prb075317}. When the device weakly couples to the leads, the TDD method will meet some serious numerical problems. To overcome those difficulties, the same group provide an exact analytical solution to the transport equations in the far from equilibrium regime for some specific voltage pulses \cite{2006prb085324}. Unfortunately, one needs to derive the new transport equations when the voltage pulses change into another shape. Obviously, it is impossible to take into account the \emph{e-e} interaction for the numerical TDD method or the analytical solutions.

 Despite the difficulty in the dynamical current calculations on the many-body systems, there are still a little works in literatures. For example, M. A. Cazalilla and J. B. Marston extends the density-matrix renormalization group approach to treat time-dependent 1D problems \cite{2002prl25}. The enhancement of the transport current through a quantum dot in the Kondo regime has been verified by their method. Except for those attempting works, the study on the
 time-dependent transport through many-body systems is far from extensive due to the computational difficulty.

In this paper, we propose a general approach based on a hierarchical equations of motion (HEOM) formalism \cite{1993prb8487,Jin08234703,2008jcp184112,2009jcp124508,2008njp093016,2009jcp164708} to characterize the time-dependent current of quantum-impurity systems. We investigate the effects of finite \emph{e-e} interaction (\emph{U}), temperature and the band width $W$ of the leads. The transient behavior of transport and time-dependent current response of interacting quantum-impurity systems are presented. Those information are relevance of experiments on quantum dots and quantum wires\cite{2002prl256403}.

The paper is organized as follows. In Sec.II we briefly review the HEOM approach, and give the common formalism for time-dependent quantum transport in quantum-impurity systems and present the results for electron current in a quantum-impurity systems driven by a external bias. Then we solve the time-dependent current using an exactly solvable model driven by a bias voltage pulse, the results are consistent with those solved by the NEGF approach in Ref.\cite{2005prb075317} in the wide-band limit(WBL). In Sec.III, with the electron-electron interaction (\emph{U}) being considered, we investigate the time-dependent current in single impurity quantum systems in and out Kondo regime at different temperatures, followed by the transport current in parallel-coupling double quantum-impurity systems. Time-dependent transport phenomena within Kondo resonance realized experimentally are also presented. In Sec.IV we give the summary of our work.

\section{GENERAL FORMALISM OF HEOM IN QUANTUM-IMPURITY SYSTEMS}
The hierarchical equations of motion approach (HEOM) is potentially useful for addressing quantum-impurity systems especially for the interacting strong correlation systems. The outstanding characterizing both equilibrium and nonequilibrium properties achieved in our previous work are referred to in Refs. \cite{Jin08234703,2012prl266403,Zhe121129}. The HEOM approach has been employed to study dynamic properties, for instance, the dynamic Coulomb blockade and dynamic Kondo memory phenomena in quantum dots \cite{2008njp093016,2009jcp164708}. It is essential to adopt appropriate truncated level to close the coupled equations. The numerical results is considered to be quantitatively accurate with increasing truncated level and converge. In Res.\cite{2012prl266403}, it has been demonstrated that the HEOM approach achieves the same level of accuracy as the latest NRG method for the prediction of various dynamical properties at equilibrium and nonequilibrium \cite{2013zheng086601}. Here, we solve the time-dependent quantum transport problem and focus on the nonequilibrium dynamics of quantum-impurity systems based on the HEOM. The localized impurities constitute the open system of primary interest, and the surrounding reservoirs of itinerant electrons are treated as environment. The truncation level of hierarchy adopts $L = 4$ to ensure numerical results presented in this letter converge quantitatively. The total Hamiltonian for the quantum-impurity systems
\begin{align}\label{ha}
   H_{T}=H_{S}+H_{B}+H_{SB}
\end{align}
where the interacting impurities
 \begin{align}\label{hs}
   H_{S}=\sum_{\mu} \epsilon_{\mu}\hat{a}^\dag_{\mu}\hat{a}_{\mu} + \frac{\mathbf{U}}{2}\sum_{\mu} n_{\mu}n_{\bar{\mu}}
 \end{align}
 the device leads treated as noninteracting electron reservoirs, in the bath interaction picture, the Hamiltonian is
 \begin{align}\label{hd}
     H_{SB}=\sum_{\mu}[f^\dag_{\mu}(t)\hat{a}_{\mu} +\hat{a}^\dag_{\mu}f_{\mu}(t)]
  \end{align}
Here, $f^\dag_{\mu}=e^{ih_{B}t}[\sum_{k}t_{\alpha k \mu}\hat{d}^\dag_{\alpha k}]e^{-ih_{B}t}$ is stochastic interactional operator and satisfies the Gauss statistics. $\hat{a}_{\mu}^\dag$ and $\hat{a}_{\mu}$ denote the creation and annihilation operators for impurity state $|\mu\rangle$ (including spin, space, \emph{etc.}), while $\hat{d}_{\alpha k}^\dag$ and $\hat{d}_{\alpha k}$ are those for the $\alpha$-reservoir state $|k\rangle$ of energy $\epsilon_{\alpha k}$. The influence of electron reservoirs on the impurities is taken into account through the hybridization functions, and the hybridization functions assume a Lorentzian form, $\Delta_{\alpha}(\w)\equiv\pi\sum_{k} t_{\alpha k}t^\ast_{\alpha k} \delta(\w-\ep_{\alpha k})=\Delta W^{2}/[2(\w-\mu_{\alpha})^{2}+W^{2}]$, where $\Delta$ is the effective impurity-lead coupling strength, $W$ is the band width, and $\mu_{\alpha}$ is the in the chemical potentials of the $\alpha$ lead. \cite{2012prl266403,2013zheng086601}.

The HEOM that governs the dynamics of open system assumes the form of \cite{Jin08234703,2012prl266403}:
\begin{align}\label{HEOM}
   \dot\rho^{(n)}_{j_1\cdots j_n} =& -\Big(i{\cal L} + \sum_{r=1}^n \gamma_{j_r}\Big)\rho^{(n)}_{j_1\cdots j_n}
     -i \sum_{j}\!     
     {\cal A}_{\bar j}\, \rho^{(n+1)}_{j_1\cdots j_nj}
\nl &
    -i \sum_{r=1}^{n}(-)^{n-r}\, {\cal C}_{j_r}\,
     \rho^{(n-1)}_{j_1\cdots j_{r-1}j_{r+1}\cdots j_n}
\end{align}
The reduced system density operator $\rho^{(0)}(t) \equiv {\rm tr}_{\B}\,\rho_{\T}(t)$ and auxiliary density operators are the basic variables. $\{\rho^{(n)}_{j_1\cdots j_n}(t); n=1,\cdots,L\}$, with $L$ denoting the terminal or truncated tier level. Here, we adopt $L=4$ is sufficient for the numerical accurate results. The Liouvillian of impurities, $\mathcal{L}\,\cdot \equiv \hbar^{-1}[H_{\s}, \cdot\,]$, may contain both \emph{e-e} interaction and time-dependent external fields.  ${\cal A}_{\bar j}$ and ${\cal C}_j$ are superoperators. The index $j \equiv (\sigma\mu m)$ corresponds to the transfer of an electron to/from ($\sigma=+/-$) the impurity state $\mu$, associated with the characteristic memory time $\gamma_m^{-1}$.

We prepare the initial total system at equilibrium, where $\mu_{\alpha}=\mu^{eq}=0$. When applying a voltage to the left(L) and the right(R)leads, the system out of equilibrium, and the time-dependent current flowing into the $\alpha$-lead $I_{\alpha}(t)$ can be presented via the first auxiliary density operator\cite{2013zheng086601}.
\begin{align}\label{hd}
    I_{\alpha}(t)=i\sum_{\mu}\mathrm{tr}_{s}[{\rho^\dag_{\alpha \mu}(t)\hat a_{\mu} -\hat a^\dag_{\mu}\rho^-_{\alpha \mu}(t)}]
  \end{align}%
Here, $\rho^\dag_{\alpha \mu}=(\rho^-_{\alpha \mu})^\dag$ is the first auxiliary density operator. So the current from $L$ to $R$ lead can be indicated $I(t)=I_{L}(t)-I_{R}(t)$. It also can be attested via conservation law of electronic flow density.
\begin{figure}
\includegraphics[width=0.9\columnwidth]{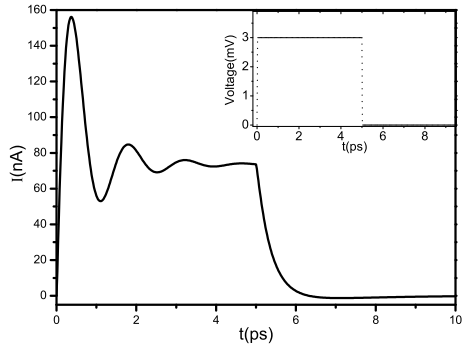}
\caption{(Solid line). The time-dependent current $I(t)$ of the  single energy level model versus time $t$ calculated by the HEOM approach under a rectangular bias voltage pulse. The parameters adopted are $V_{L}(t)=-V_{R}(t)= 3\mathrm{mV}$, $\epsilon_{\mathrm{d}} = 0.1\mathrm{meV}$,$W = 16\mathrm{meV}$, $\Delta=0.5\mathrm{meV}$. For the case of resonance tunneling, the current is almost independent of the temperature $T$, thus the value of $T$ is not shown here.
The inset shows the rectangular bias voltage pulses applied into the left lead.} \label{fig1}.
\end{figure}

As a test, we consider a single energy level model in response to a rectangular bias voltage pulses as shown in the inset of Figure 2(a), which has been studied in Ref.\cite{2005prb075317} by the NEGF method. For comparison we neglect the electron-electron interaction $U$ in HEOM approach like the NEGF dose.

Figure~\ref{fig1} depicts the calculated transport current $I(t)$ by the HEOM approach and up to the converged tier level($L=4$), where, $\Delta$ is the effective impurity-reservoir coupling strength and $W$ is the reservoir band width. The parameters adopted are $V_{L}(t)=-V_{R}(t)= 3\mathrm{mV}$, $\epsilon_{\mathrm{d}} = 0.1\mathrm{meV}$,$W = 16\mathrm{meV}$, $\Delta=0.5\mathrm{meV}$. For the case of resonance tunneling, the current is almost independent of the temperature $T$, thus the value of $T$ is not shown here. At $t =0$, a bias pulse (dashed curve) suddenly increases energies in the leads by $V_{L}(t)=-V_{R}(t)=3\mathrm{mV}$ (see inset of Figure~\ref{fig1}). At $t= 5\mathrm{ps}$, before the current has settled to a new steady value, the pulse ends and the current rapid decays back to zero. As show in the Figure 2.(a) in Ref.\cite{2005prb075317}, in the time $0<t<5\mathrm{ps}$, the current exhibits oscillation around its steady state value(Rabi oscillation) \cite{1993prb8487}. This phenomenon may be observed experimentally in the dc current by varying the voltage pulses applied to the structure. The results of time-dependent current curve recovers those obtained by the approach of NEGF within the WBL approximation \cite{2005prb075317}, except the WBL and as well as other approximations are totally necessary in the HEOM.

\section{CURRENT DRIVEN BY VOLTAGE PULSE IN QUANTUM-IMPURITY SYSTEMS}

Now, we consider the oscillation of dynamical current in the Kondo regime, which is much more important than above single-level resonance tunneling. There is no energy level near the Fermi energy or within the transport window in this case, thus neither large current nor current oscillation can be observed without the Kondo effect. As we know, the Kondo resonance at low temperature will induce large current. In this case, a new kind of current oscillation behavior different from that in Figure~\ref{fig1} is expected. We will study the oscillation of dynamical current for such two cases: 1) single quantum-impurity system; and 2) parallel-coupling double quantum-impurity system. For the former case, we will detailedly investigate the dependence of the current oscillation on various factors, such as the form of the bias voltage, temperature and band width of the leads.

\subsection{single quantum-impurity systems}

We first consider the single quantum-impurity systems, the Hamiltonian of the impurity can be written as
 \begin{align}\label{hs}
   H_{single}=\epsilon_{\uparrow}\hat{a}^\dag_{\uparrow}\hat{a}_{\uparrow} +\epsilon_{\downarrow}\hat{a}^\dag_{\downarrow}\hat{a}_{\downarrow} + U n_{\uparrow}n_{\downarrow}
 \end{align}
Here, $\hat{a}_{\uparrow\downarrow}^\dag$ and $\hat{a}_{\uparrow\downarrow}$ denote the creation and annihilation operators for spin up and down of the impurity, $U$ is the electron-electron interaction.

\begin{figure}
\includegraphics[width=0.85\columnwidth]{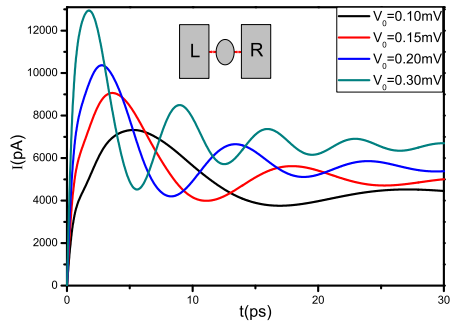}
\caption{(Color online). The dynamical current $I(t)$ of single quantum-impurity systems subject to an step voltage with constant value $V_{0}$. The parameters adopted are $K_{B}T=0.015\mathrm{meV}$, $\Delta=0.2\mathrm{meV}$, $W=2.0\mathrm{meV}$, $U=\mathrm{meV}$ and $\epsilon_{\uparrow}=\epsilon_{\downarrow}=-1\mathrm{meV}$. }
\label{fig2}
\end{figure}

Figure~\ref{fig2} depicts the current $I(t)$ characteristics of a single quantum-impurity systems which possesses the electron-hole symmetry ($\epsilon_{\uparrow}=\epsilon_{\downarrow}=-U/2$) subject to various forms of the time-dependent step voltage.
\begin{align}\label{hd}
   V(t)=\left\{
   \begin{array}{ll}
   0 \;\,\qquad(t<0)& \\
   V_{0} \qquad(t\geq 0) &
   \end{array}
    \right.
  \end{align}%
$V_{0}$ is the step voltage values, $V_{0}=0.10\mathrm{mV}$, $0.15\mathrm{mV}$, $0.20\mathrm{mV}$, $0.30\mathrm{mV}$ respectively. As shown in Figure~\ref{fig1}, when the voltage pulse applied to the leads, the current flowing through the device engenders. After the current rapidly increases to a maximal value, regular Rabi oscillations emerge, which results from the temporal coherence of electrons tunneling through the quantum impurity in response to the abrupt change of bias voltage. We find that the form of the oscillations depends strongly on the size of the applied step voltage. With the increase of the voltage, the time-dependent current become stronger and the amplitude of oscillations intensify much. For example, the maximal amplitude of current$I(t)$ is only $7000\mathrm{pA}$ at the voltage $V_{L}=-V_{R}=0.10\mathrm{mV}$, while it reaches to $13000\mathrm{pA}$ at the voltage $V_{L}=-V_{R}=0.30\mathrm{mV}$, meanwhile the frequency of the oscillations increases. On the other side, all of the current values will arrive at different stable values at $t>30\mathrm{ps}$ independent of the form of the voltage pulse. Those current values obviously corresponds to the steady-state current.

\begin{figure}
\begin{minipage}[t]{0.5\linewidth}
\raggedleft
\includegraphics[width=2.0in]{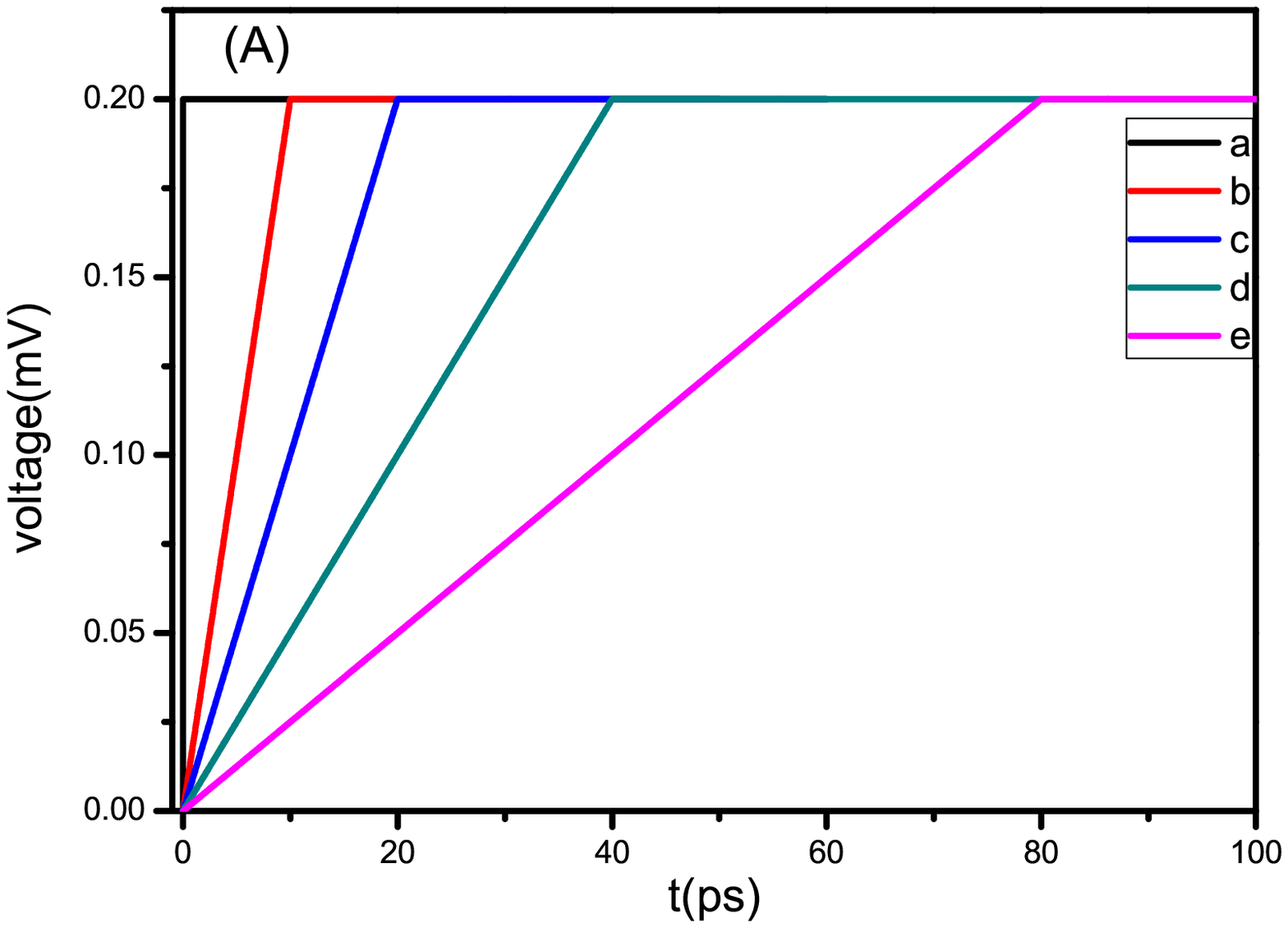}

\label{fig3(A)}
\end{minipage}%
\begin{minipage}[t]{0.5\linewidth}
\raggedleft
\includegraphics[width=2.0in]{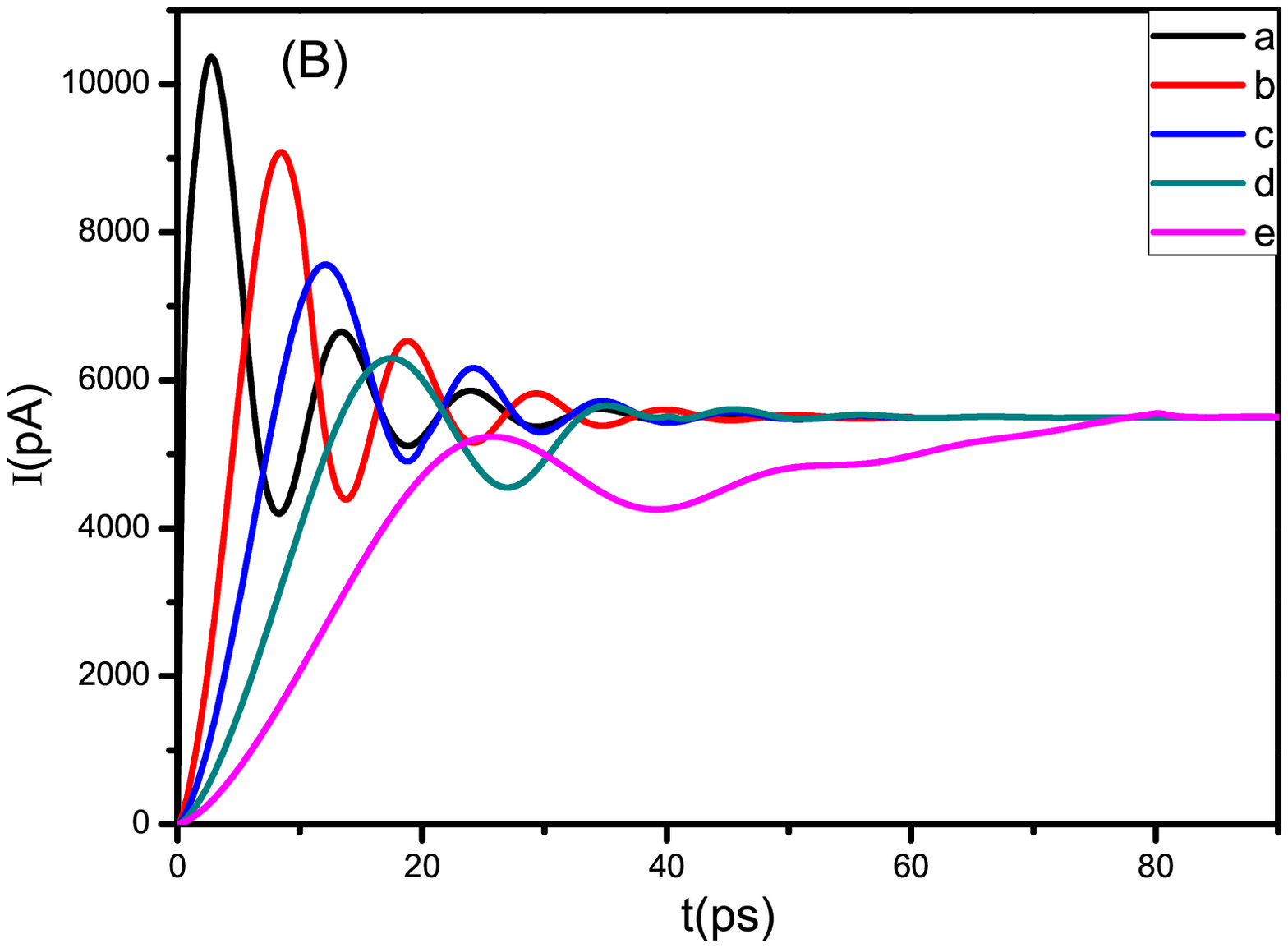}

\label{fig3(B)}
\end{minipage}
\caption{(Color online) (A) Linear-increasing voltages $V(t)$ applied on the leads of the quantum-impurity system, which increase to $V_m(t)=0.20\mathrm{mV}$ through different intervals of time $t_{c}$: $a$ ($t_{c}=0$, step-increasing) , $b$ ($t_{c}=10\mathrm{ps}$), $c$ ($t_{c}=20\mathrm{ps}$), $d$ ($t_{c}=40\mathrm{ps}$), and $e$ ($t_{c}=80\mathrm{ps}$). (B) Time-dependent current $I(t)$ for different linear-increasing voltages $V(t)$ as described in Fig3(A). The other parameters are $K_{B}T=0.015\mathrm{meV}$, $\Delta=0.2\mathrm{meV}$, $W=2.0\mathrm{meV}$, $U=2\mathrm{meV}$ and $\epsilon_{\uparrow}=\epsilon_{\downarrow}=-1\mathrm{meV}$. }
\end{figure}
To check the details of the influence of the voltage on the behavior of current$I(t)$, we alter the step-increasing voltages into linear-increasing ones. As show in Figure 3(A), the voltages linearly increase to $V_m(t)=0.20\mathrm{mV}$ through different intervals of time $t_{c}$. $a$ is the step-increasing one for comparison, $b$, $c$, $d$, and $e$ are linear-increasing ones with $t_{c}=10\mathrm{ps}$, $20\mathrm{ps}$, $40\mathrm{ps}$, and $80\mathrm{ps}$, respectively. Figure 3(B) plots the time-dependent current $I(t)$ for above linear-increasing voltages under the same temperature $T$ and band width $W$. The other parameters are $K_{B}T=0.015\mathrm{meV}$, $\Delta=0.2\mathrm{meV}$, $W=2.0\mathrm{meV}$, $U=2\mathrm{meV}$ and $\epsilon_{\uparrow}=\epsilon_{\downarrow}=-1\mathrm{meV}$. It is interesting to see that the currents under the linear-increasing voltages remain  oscillations, which decrease with the increase of time and then vanish after enough long time. The maximal amplitudes of the current  reduce with the increase of $t_{c}$, for example, the maximal value of the current gets up to $9000\mathrm{pA}$ for $t_{c}=10\mathrm{ps}$, while it only reaches $5000\mathrm{pA}$ for $t_{c}=80\mathrm{ps}$.

It has been demonstrated that the oscillations of the current mainly results from the temporal coherence of electrons tunneling through the device, which can be manipulated by the way of the applied voltage as shown in Figure 3. For $t_{c}=10\mathrm{ps}$, the oscillation of $I(t)$ is fast (with a short period) and dramatic (with a large amplitude), since the carriers have insufficient time to redistribute to catch up the variation of voltage \cite{2013zheng086601}. In contrast, when $t_{c}$ increase to $80\mathrm{ps}$, the oscillation of $I(t)$ becomes very slow and insignificant
due to the longer relaxation time of the carriers.

\begin{figure}
\includegraphics[width=0.95\columnwidth]{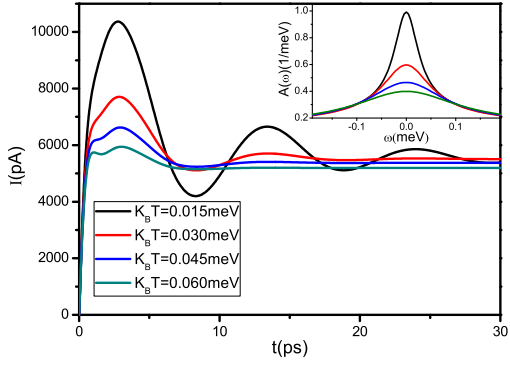}
\caption{(Color online). The $I(t)$ - $t$ curve of the symmetric quantum-impurity system with different temperatures $T$. The inset shows the spectral function of the symmetric quantum-impurity system. The parameters adopted are $V_{L}=-V_{R}=0.20\mathrm{mV}$, $\Delta=0.2\mathrm{meV}$, $W=2.0\mathrm{meV}$, $U=2\mathrm{meV}$ and $\epsilon_{\uparrow}=\epsilon_{\downarrow}=-1\mathrm{meV}$.}
\label{fig4}
\end{figure}

To further understand those effects of the transport current, we then vary the temperature $T$, the band width $W$ and the coupling strength $U$ and summarize the results in Figure~\ref{fig4} to 6. In Figure~\ref{fig4}, we show the change of the $I(t)$-$t$ curve with the temperature in the Kondo regime ($T<T_{K}$). As shown in the figure, the oscillation of $I(t)$ is distinct with large amplitude at very low temperature such as $K_{B}T=0.015\mathrm{meV}$. With the increase of $T$, the oscillation will be suppressed gradually, for example, it changes to a little wobble around the steady current at $K_{B}T=0.045\mathrm{meV}$. At $K_{B}T=0.060\mathrm{meV}$, the oscillation almost dies away and $I(t)$ reaches its stable value very quickly. In the insert of Figure~\ref{fig4}, we show the spectral functions of the system without the bias voltage around $\omega =0$ corresponding to above four temperatures. As shown in the figure, the  height of the  Kondo peak is reduced with the increase of temperature, indicating the suppression of the Kondo effects. Once the bias voltages apply to the leads of the device, each of the Kondo resonance peak at $\omega =0$ is splitted into two peaks at $\omega=\pm eV$ ($V$ is the stable value of the voltage). The temperature plays the restraining role in the  dynamic transport of the quantum-impurity systems. Summarizing Figure~\ref{fig4} and its inserting figure, one can conclude that the temperature suppresses the oscillation of dynamic current through its suppression to the Kondo effects. The temperature-dependent current oscillation shown in Figure~\ref{fig4} is totally different from temperature-independent case shown in Figure~\ref{fig1}.

\begin{figure}
\includegraphics[width=0.95\columnwidth]{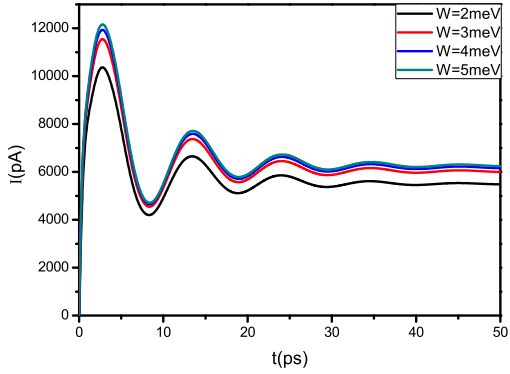}
\caption{(Color online). The $I(t)$ - $t$ curve of the symmetric quantum-impurity system with different band width $W$. The parameters adopted are $K_{B}T=0.015\mathrm{meV}$, $V_{L}=-V_{R}=0.20\mathrm{mV}$, $K_{B}T=0.015\mathrm{meV}$, $\Delta=0.2\mathrm{meV}$, $U=2\mathrm{meV}$ and $\epsilon_{\uparrow}=\epsilon_{\downarrow}=-1\mathrm{meV}$.}
\label{fig5}
\end{figure}

We then elucidate the influence of the finite band width $W$ on the current oscillation, which is hard to treat by NEGF approach within the WBL approximation. The characteristics of dynamic $I(t)$ corresponding to different band widths are shown in Figure ~\ref{fig5}. It can be seen that the larger band width $W$ principally leads to larger amplitude of current oscillation as well as larger value of steady-state current. That effect mainly results from the $W$-enhancement of the capacitive contributions from the accumulation and depletion of electrons layering on either side of the device leads. On the other hand, as shown in Figure ~\ref{fig5}, the frequency of the oscillation is almost independent from the band width $W$.

\begin{figure}
\begin{minipage}[t]{0.5\linewidth}
\raggedleft
\includegraphics[width=2.0in]{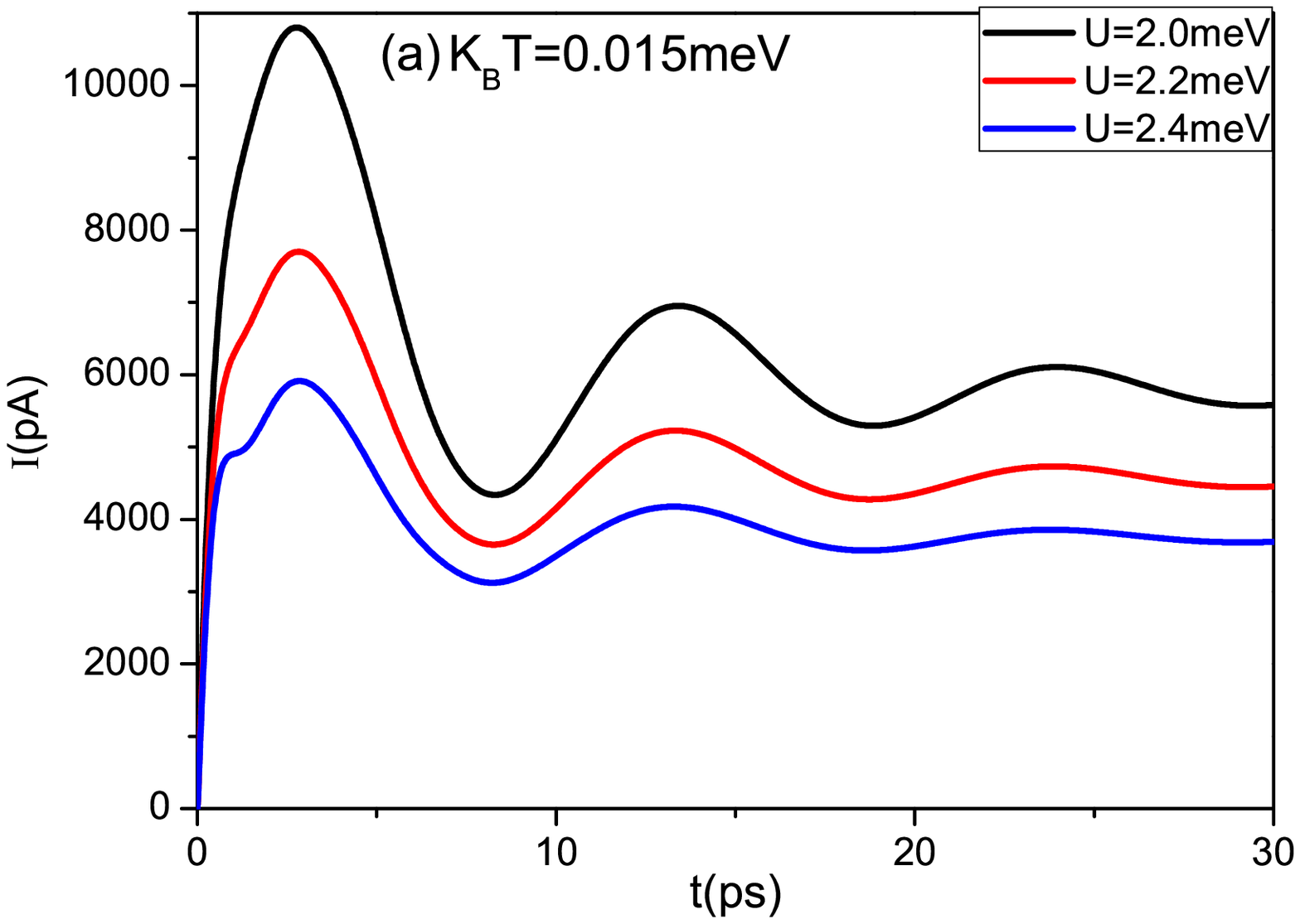}

\label{fig6a}
\end{minipage}%
\begin{minipage}[t]{0.5\linewidth}
\raggedleft
\includegraphics[width=2.0in]{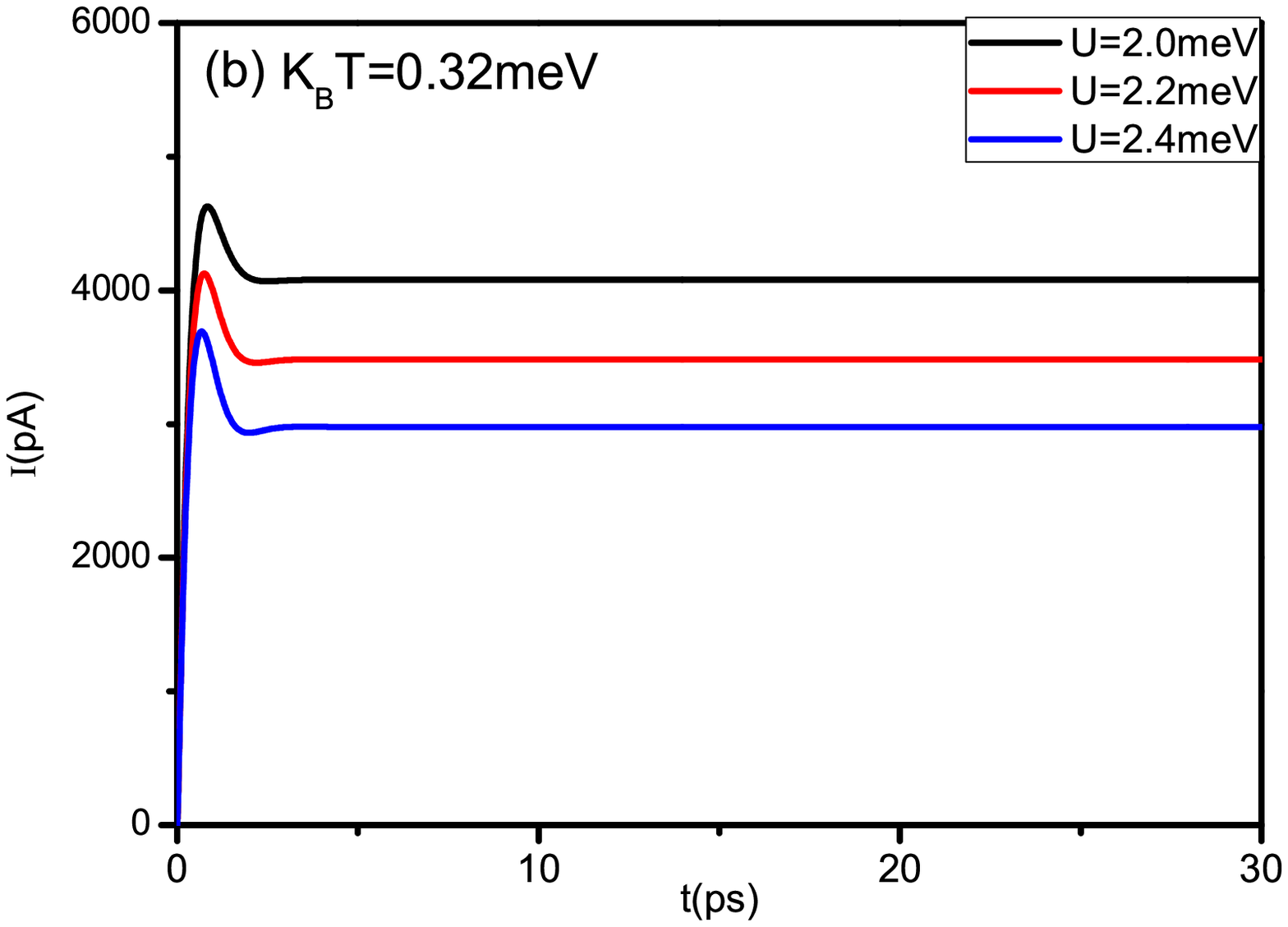}

\label{fig6b}
\end{minipage}
\caption{(Color online) The $I(t)$-$t$ curve of the symmetric quantum-impurity system with different \emph{e-e} interaction $U$. (A) in the Kondo regime($T<T_{K}$), the parameters adopted are $V_{L}=-V_{R}=0.20\mathrm{mV}$, $W=2.0\mathrm{meV}$, $K_{B}T=0.015\mathrm{meV}$, $\Delta=0.2\mathrm{meV}$, and $\epsilon_{\uparrow}=\epsilon_{\downarrow}=-U/2$. (B) out of the Kondo regime($T>T_{K}$), the parameters adopted are $V_{L}=-V_{R}=0.20\mathrm{mV}$, $W=2.0\mathrm{meV}$, $K_{B}T=0.32\mathrm{meV}$, $\Delta=0.2\mathrm{meV}$, and $\epsilon_{\uparrow}=\epsilon_{\downarrow}=-U/2$.}
\end{figure}

In the end of this subsection, we investigate the effect of \emph{e-e} interaction $U$ on the current oscillation. In our calculations, the electron-hole symmetry is kept, i.e. the on-site energy of each dot is chosen as $\epsilon_{\uparrow}=\epsilon_{\downarrow}=-U/2$. Figure 6 shows the results for both cases of $T<T_{K}$ [Figure 6(A)] and $T>T_{K}$ [Figure 6(B)]. Overall speaking, the on-site \emph{e-e} interaction will induce the localization of the carriers, which make the steady current decrease with the increase of $U$, as shown both in Figure 6(A) and 6(B). In the Kondo regime, the amplitude of the current oscillation decreases with the increase of $U$, but the frequency keeps almost unchanged. The mechanism can be understood as follows: According to the analytical expression for Kondo temperature $T_{K}=\sqrt{\frac{U\Delta}{2}}e^{-\pi U/8\Delta+\pi\Delta/2U}$ \cite{1993cambrige}, $T_{K}$ decreases with the increase of $U$. Since the temperature is fixed (at $K_{B}T=0.015\mathrm{meV}$), a larger $U$ will induce a smaller distance between $T$ and $T_{K}$. As a consequence, the current oscillation will take place more dramatically for smaller $U$, for example, the maximal amplitude of the oscillation increases from $4600\mathrm{pA}$ at $U=2.4$ meV  to $11000\mathrm{pA}$ at $U=2.0$ meV. Summarizing Figure 2-6, one can conclude that the frequency of the current oscillation is strongly dependent on the bias voltage $V(t)$ but almost independent of $T$, $W$ and $U$ as well.

\subsection{double quantum-impurity systems}

Double quantum-impurity systems, also considered as artificial molecules, are very important for the study of many novel phenomena such as two-channel Kondo effects and non-Fermi liquid behavior. Moreover, those structure are convenient to realize the solid state quantum bits as reported in literatures \cite{2000pla271,1995prl705,1998prl4032}. Here, we just focus on the oscillation of the dynamical current in the parallel-coupling double quantum-impurity systems, with each of the impurity keeping single occupation. The Hamiltonian for such sub-systems can be written as,
\begin{align}\label{hs}
   H_{double}=\epsilon_{1\uparrow}\hat{a}^\dag_{1\uparrow}\hat{a}_{1\uparrow} +\epsilon_{1\downarrow}\hat{a}^\dag_{1\downarrow}\hat{a}_{1\downarrow} + U_{1} n_{1\uparrow}n_{1\downarrow}
   \nl +\epsilon_{2\uparrow}\hat{a}^\dag_{2\uparrow}\hat{a}_{2\uparrow}
   +\epsilon_{2\downarrow}\hat{a}^\dag_{2\downarrow}\hat{a}_{2\downarrow} + U_{2} n_{2\uparrow}n_{2\downarrow}
  \nl +\Gamma_{12}\hat{a}^\dag_{1\uparrow}\hat{a}_{2\uparrow}
   +\Gamma_{12}\hat{a}^\dag_{1\downarrow}\hat{a}_{2\downarrow}+\Gamma_{21}\hat{a}^\dag_{2\uparrow}\hat{a}_{1\uparrow}
   +\Gamma_{21}\hat{a}^\dag_{2\downarrow}\hat{a}_{1\downarrow}
 \end{align}
Here, $\hat{a}_{1\uparrow1\downarrow}^\dag$ and $\hat{a}_{1\uparrow1\downarrow}$ denote the creation and annihilation operators for spin up and down of the impurity one, $\hat{a}_{2\uparrow2\downarrow}^\dag$ and $\hat{a}_{2\uparrow2\downarrow}$ denote the creation and annihilation operators for spin up and down of the impurity two, $U_{1}$ and $U_{2}$ are respectively the \emph{e-e} interaction of impurities. $\Gamma_{12}$ and $\Gamma_{21}$ are inter-impurity couplings between the two impurities, being set $\Gamma_{12}=\Gamma_{21}=\Gamma$ in our calculations.
\begin{figure}
\includegraphics[width=0.95\columnwidth]{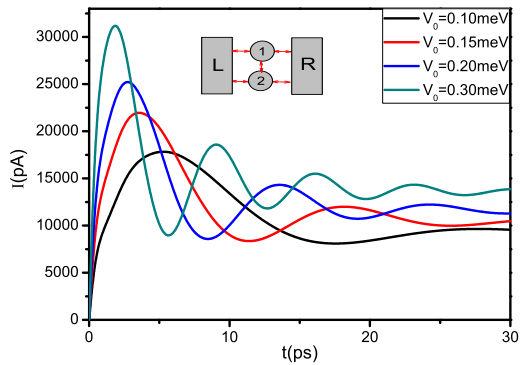}
\caption{(Color online). The $I(t)$ - $t$ curve of the parallel-coupling double quantum-impurity system subject to a step voltage with constant value $V_{0}$. The parameters adopted are $K_{B}T=0.015\mathrm{meV}$, $\Delta=0.2\mathrm{meV}$, $W=2.0\mathrm{meV}$, $U_{1}=U_{2}=2\mathrm{meV}$, $\epsilon_{1\uparrow}=\epsilon_{1\downarrow}=-1\mathrm{meV}$, $\epsilon_{2\uparrow}=\epsilon_{2\downarrow}=-1\mathrm{meV}$, $\Gamma=0.1\mathrm{meV}$.}
\label{fig7}
\end{figure}

Figure~\ref{fig7} depicts the dynamical current $I(t)$ of a parallel-coupling double quantum-impurity system subject to a step voltage in the Kondo regime. In our calculations, each of the impurity holds the electron-hole symmetry, with the same parameters as the single quantum-impurity system as shown in Figure~\ref{fig2}. For comparison, the constant values of the step voltage are also chosen the same as those in Figure~\ref{fig2}, \emph{i.e.} $V_{0}=0.10\mathrm{mV}$, $0.15\mathrm{mV}$, $0.20\mathrm{mV}$ and $0.30\mathrm{mV}$. In Figure~\ref{fig7}, we take the tunneling coupling between the two impurities as a weak one $\Gamma=0.1\mathrm{meV}$. As reported in our previous work \cite{2012prl266403}, this weak coupling maintains the Kondo singlet of each impurity, which will double the transport channels. That point has been clearly elucidated in Figure~\ref{fig7} by the following two characteristics: 1) the steady current of the double impurity system (about $10000\sim15000$ pA) is almost twice the single one (about $5000\sim7000$ pA, see Figure~\ref{fig2}); and 2) the oscillation of $I(t)$ of the double impurity system shows similar behaviors to the single one, but with two-to-three times larger amplitudes. For example, at the voltage $V_{L}=-V_{R}=0.20\mathrm{mV}$, the maximal value of current $I(t)$ is about $10200\mathrm{pA}$ (see Figure~\ref{fig2}) in single quantum-impurity systems, while it rises to $25000\mathrm{pA}$ in parallel-coupling double quantum-impurity systems as shown in Figure~\ref{fig7}.

\begin{figure}
\includegraphics[width=0.95\columnwidth]{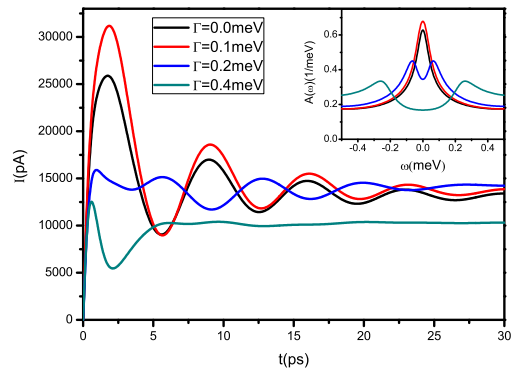}
\caption{(Color online). Time-dependent current $I(t)$ - $t$ curve of the double quantum-impurity system with different tunneling coupling between the two impurities $\Gamma$. The inset shows the spectral function of the symmetric parallel-coupling double quantum-impurity systems. The parameters adopted are $K_{B}T=0.015\mathrm{meV}$, $\Delta=0.2\mathrm{meV}$, $W=2.0\mathrm{meV}$, $V_{L}=-V_{R}=0.3\mathrm{mV}$,  $U_{1}=U_{2}=2\mathrm{meV}$, $\epsilon_{1\uparrow}=\epsilon_{1\downarrow}=-1\mathrm{meV}$, $\epsilon_{2\uparrow}=\epsilon_{2\downarrow}=-1\mathrm{meV}$.}
\label{fig8}
\end{figure}

One interesting issue in double quantum-impurity systems is the quantum phase transition induced by the inter-impurity coupling $\Gamma$, from the degenerate Kondo singlet states of individual impurity to singlet spin states formed between two impurities\cite{2012prl266403}. Let us illustrate this picture intuitively from the point of view of the dynamical current responding to the same bias voltage but different inter-impurity couplings. The calculated results are shown in Figure~\ref{fig8}, where the parameters are chosen as $K_{B}T=0.015\mathrm{meV}$, $\Delta=0.2\mathrm{meV}$, $W=2.0\mathrm{meV}$, $V_{L}=-V_{R}=0.3\mathrm{mV}$,  $U_{1}=U_{2}=2\mathrm{meV}$, $\epsilon_{1\uparrow}=\epsilon_{1\downarrow}=-1\mathrm{meV}$, $\epsilon_{2\uparrow}=\epsilon_{2\downarrow}=-1\mathrm{meV}$. Together with the time-dependent current $I(t)$ - $t$ curves, we also show the total equilibrium spectral function of the system ($V_{L}=V_{R}=0$) in the inset of Figure~\ref{fig8}. By referring the figure, one can see that with the increase of $\Gamma$ the oscillation of the dynamical current changes dramatically, with some features of the quantum phase transition clearly shown. In the regime of small $\Gamma$, the direct first-order coupling ($\Gamma$) is much stronger than the induced second-order antiferromagnetic spin coupling ($J=4\Gamma^2/U$) between two impurities. As a consequence, the ground sate of the systems is the degenerate Kondo singlet states of individual impurity, with the equilibrium spectral function showing single peak at $\omega=0$. The behavior of the current oscillation in this case is analogous to the single impurity one, as already confirmed in Figure~\ref{fig7}. Besides, a new feature called `$\Gamma-$enhanced Kondo effect' can be seen in Figure~\ref{fig8}, which shows itself by the increased amplitude but the same frequency in the oscillation of the dynamical current (compare the $I-t$ curves of $\Gamma=0$ and $0.1$ meV). We will publish our detail study on the $\Gamma-$enhanced Kondo effect elsewhere. In the regime of large $\Gamma$, the spin-spin coupling $J$ dominates which locks the ground state as spin-singlet states between two impurities. The Kondo peak of the  equilibrium spectral function at $\omega=0$ thus disappears, and consequently the current oscillation is suppressed, as shown in Figure~\ref{fig8} (see the $I-t$ curve of $\Gamma=0.4$ meV).

In the regime of medium $\Gamma$, the crossover of the phase transition between the above two ground state takes place. By referring Figure~\ref{fig8}, we can find two interesting features of the current oscillation: 1) although the peak of density of state at $\omega=0$ has already changed to a dip, the double peaks ($\sim \pm J=0.8$ meV) in the transition window ($-0.3-+0.3$ meV) still cause the oscillations of current with smaller amplitude; and 2) the frequency of the current oscillation for the double-peak and single-peak states is almost the same, but the phase is exactly reversed (compare the $I-t$ curves of $\Gamma=0.1$ and $0.2$ meV for details).

\section{SUMMARY}

In summary, we have investigated the time-dependent transport properties of single and double  quantum-impurity systems based on the hierarchical equations of motion. In the Kondo regime, the dynamical current in both cases is found oscillating due to the temporal coherence of electrons tunneling through the device, which shares the same mechanism as the single-level resonance without \emph{e-e} interactions but shows some different characteristics.

For single quantum-impurity systems, the temperature plays an inhibitory action to the oscillations of dynamic current through its suppression to the Kondo effects. The amplitude of the current oscillations is found being enhanced by the band width $W$ of the leads but attenuated by the \emph{e-e} interaction $U$ in the Kondo regime. The reason for the latter is that the \emph{e-e} interaction will partly weaken the Kondo effect at the fixed temperature $T$ by means of inducing a smaller distance between $T$ and $T_{K}$. We find the steady-state current decreases with the increase of $U$ for both cases of $T<T_{K}$ and $T>T_{K}$. On the other hand, the frequency of the current oscillation is found almost independent of $T$, $W$ and $U$.

For parallel-coupling double quantum-impurity systems, the steady current is approximately twice of the single one and the oscillation of $I(t)$ shows similar behaviors to the single one, but with two-to-three times larger amplitudes. With the increase of the inter-impurity coupling $\Gamma$, the system undergos a quantum phase transition process from the degenerate Kondo singlet states of individual impurity to singlet spin states formed between two impurities. Reflected in the dynamical current, we find that at low $\Gamma$ the oscillation of the current exhibits enhanced characters due to the `$\Gamma-$enhanced Kondo effect', while the oscillation will be subdued at the other limit (with large $\Gamma$). Those characteristics may be observed in experiments.

%
%
The support from the NFS of China (No.\,11374363)and the the Research Funds of Renmin
University of China (Grant No. 11XNJ026) is gratefully appreciated.
%




\end{document}